\documentclass[conference]{IEEEtran}

\usepackage{algorithm}
\usepackage{algorithmic}
\usepackage{amsmath}
\usepackage{amssymb}
\usepackage{amsfonts}
\usepackage{graphicx}
\usepackage{graphics}
\usepackage{setspace}
\usepackage{epsfig}
\usepackage{subfigure}
\usepackage{psfrag}
\usepackage{cite}
\usepackage{latexsym}
\usepackage{url}
\usepackage{color}
\usepackage{multirow}
\DeclareMathOperator*{\argmax}{arg\,max}

\PassOptionsToPackage{bookmarks={false}}{hyperref}

\begin{document}
\title{Capacity of UAV-Enabled Multicast Channel: Joint Trajectory Design and Power Allocation
}
\author{Yundi Wu$^1$, Jie Xu$^2$, Ling Qiu$^1$, and Rui Zhang$^3$\\
$^1$School of Information Science and Technology, University of Science and Technology of China\\
$^2$School of Information Engineering, Guangdong University of Technology\\
$^3$Department of Electrical and Computer Engineering, National University of Singapore\\
E-mail:~wyd57@mail.ustc.edu.cn,~jiexu@gdut.edu.cn,~lqiu@ustc.edu.cn,~elezhang@nus.edu.sg

}

\maketitle

\begin{abstract}
This paper studies an unmanned aerial vehicle (UAV)-enabled multicast channel, in which a UAV serves as a mobile transmitter to deliver common information to a set of $K$ ground users. We aim to characterize the capacity of this channel over a finite UAV mission/communication period, subject to its maximum speed constraint and an average transmit power constraint. To achieve the capacity, the UAV should use a sufficiently long code that spans over its whole mission/communication period. Accordingly, the multicast channel capacity is achieved via maximizing the minimum achievable time-averaged rates of the $K$ users, by jointly optimizing the UAV's trajectory and transmit power allocation over time. However, this problem is non-convex and difficult to be solved optimally. To tackle this problem, we first consider a relaxed problem by ignoring the maximum UAV speed constraint, and obtain its globally optimal solution via the Lagrange dual method. The optimal solution reveals that the UAV should hover above a finite number of ground locations, with the optimal hovering duration and transmit power at each location. Next, based on such a multi-location-hovering solution, we present a {\emph{successive hover-and-fly}} trajectory design and obtain the corresponding optimal transmit power allocation for the case with the maximum UAV speed constraint. Numerical results show that our proposed joint UAV trajectory and transmit power optimization significantly improves the achievable rate of the UAV-enabled multicast channel, and also greatly outperforms the conventional multicast channel with a fixed-location transmitter.
\end{abstract}
\begin{keywords}
Unmanned aerial vehicle (UAV), multicast channel, capacity, trajectory design, power allocation.
\end{keywords}

\newtheorem{definition}{\underline{Definition}}[section]
\newtheorem{fact}{Fact}
\newtheorem{assumption}{Assumption}
\newtheorem{theorem}{\underline{Theorem}}[section]
\newtheorem{lemma}{\underline{Lemma}}[section]
\newtheorem{corollary}{\underline{Corollary}}[section]
\newtheorem{proposition}{\underline{Proposition}}[section]
\newtheorem{example}{\underline{Example}}[section]
\newtheorem{remark}{\underline{Remark}}[section]

\newcommand{\mv}[1]{\mbox{\boldmath{$ #1 $}}}
\setlength\abovedisplayskip{6pt}
\setlength\belowdisplayskip{6pt}

\section{Introduction}
Unmanned aerial vehicle (UAV)-enabled wireless communications have recently attracted a lot of interests from both academia and industry, as UAVs can be utilized as aerial communication platforms (such as base station (BSs) and relays) to provide wireless access for mobile subscribers on the ground (see, e.g., \cite{ZengZhangLim2016} and the references therein). Compared to conventional terrestrial wireless communication systems, UAV-enabled systems possess the following main advantages. First, different from the fixed wireless infrastructures on the ground, UAVs can be deployed swiftly on-demand, and thus are suitable for unexpected or urgent scenarios (e.g., in natural disasters). Second, the air-to-ground as well as ground-to-air wireless communication links between UAVs and ground users are more likely to be line-of-sight (LoS); therefore, UAV-enabled wireless communications in general have better channel conditions than terrestrial channels under the same link distance. Last but not least, UAVs have the fully controllable mobility, and thus can adjust their locations over time to shorten the distances to intended ground users for enhancing the communication performance.

In the literature, there have been a handful of works investigating the UAV-enabled wireless communication systems (see, e.g., \cite{LyuZengZhang2017,BorEl2016,Mozaffari2016,Wu2017,ZengZhangLim2016-2,WuXuZhang2017}). Specifically, in \cite{LyuZengZhang2017,BorEl2016,Mozaffari2016}, UAVs are used as aerial BSs to provide wireless coverage for ground users, in which the UAVs' placement or their quasi-static hovering locations over a certain period of time have been optimized to improve the coverage performance and/or reduce the system cost. In \cite{Wu2017,ZengZhangLim2016-2,WuXuZhang2017}, the authors exploit the UAVs' fully controllable mobility via trajectory optimization to improve the communication rates of UAV-enabled wireless networks. Furthermore, trajectory optimization has also been investigated in UAV-enabled wireless power transfer (WPT) systems \cite{Xu2017,XuZengZhang2017} and wireless powered communication networks (WPCN) \cite{XieXuZhang2018} to improve the energy and communication performance. Motivated by the above works, in this paper we study a new UAV-enabled information multicasting system, in which a UAV serves as a mobile transmitter to send common information to a set of ground users. We aim to address the open question of how to optimize the UAV's trajectory jointly with its transmit power allocation to achieve the capacity of the UAV-enabled multicast channel.

Multicast channels with fixed transmitters have been extensively investigated in wireless communications, due to its wide applications in e.g. video streaming and information dissemination. Conventionally, multi-antenna techniques have been adopted as a promising solution to improve the communication rate of a multicast channel. For instance, \cite{Jindal2006} studies the capacity limits of multi-antenna multicast channels, and \cite{Sidiropoulos2006,Xiang2013,Wu2013} develop practical transmit beamforming designs to improve the achievable rate to approach the capacity. However, in a multicast channel, as all user receivers need to successfully decode the common information from the transmitter, its capacity is fundamentally limited by the user with the worst channel condition, even with multi-antenna beamforming applied at the transmitter. In particular, if one user is much farther away from the transmitter than the others, the capacity of the multicast channel will be fundamentally constrained by this user.

To overcome such limitations, in this paper, we study a new UAV-enabled multicast channel, in which a UAV serves as a {\it mobile} transmitter to deliver common information to a set of $K$ ground users. Different from the conventional multicast channel with a fixed transmitter, the mobile transmitter (or UAV) can adjust its location over time to improve the wireless channels to different users at different locations. This overcomes the bottleneck user issue in conventional multicast channels with a fixed transmitter and thus significantly enhances the capacity. To reveal the fundamental limit of the UAV-enabled multicast channel, we characterize its capacity over a finite UAV mission/communication period, subject to its maximum speed constraint and an average transmit power constraint. To achieve the capacity, the UAV should use a capacity-achieving code that is sufficiently long to span over the whole mission/communication period. Therefore, the multicast channel capacity is obtained via solving the problem of maximizing the minimum achievable time-averaged rates of the $K$ users, by jointly optimizing the UAV's trajectory and transmit power allocation over time.

However, the joint trajectory design and power control problem is difficult to be solved optimally due to its non-convexity. To tackle this problem, we first consider a relaxed problem by ignoring the maximum UAV speed constraint. We show that the relaxed problem satisfies the so-called time-sharing condition in \cite{YuLui2006}, and thus can be optimally solved via the Lagrange dual method. The optimal solution to the relaxed problem reveals that the UAV should hover above a finite number of ground locations during the whole mission/communication period, with the hovering duration and transmit power at each location optimized. Next, for the problem with the maximum UAV speed constraint considered, we present an efficient successive hover-and-fly trajectory design based on the optimal multi-location-hovering solution to the relaxed problem, and obtain the corresponding optimal transmit power allocation. Finally, numerical results are provided to show the benefit of our proposed joint UAV trajectory and transmit power optimization in the UAV-enabled multicast channel, as compared to the conventional multicast channel with a fixed-location transmitter as well as other benchmark schemes.

To our best knowledge, there is only one prior work in the literature \cite{ZengXu2017} that considers a similar UAV-enabled information multicasting system, in which a UAV is dispatched to disseminate a common large file to a number of distributed ground users by using practical random linear network coding (RLNC). The objective in \cite{ZengXu2017} is to  minimize the mission completion time via optimizing the UAV trajectory, while ensuring that each ground user is able to successfully recover the file with a high probability. By contrast, this paper characterizes the fundamental capacity of a UAV-enabled multicast channel, by considering the joint optimization of the UAV trajectory design and transmit power allocation. This problem is new and has not been studied before.

\section{System Model and Problem Formulation}

We consider a UAV-enabled multicast channel, in which a UAV serves as a mobile transmitter to send common information to $K > 1$ randomly located users on the ground. Let $\mathcal{K} \triangleq \{ 1,...,K \}$ denote the set of ground users. We consider a three-dimensional (3D) Cartesian coordinate system, and suppose that each user $k \in \mathcal{K}$ has a fixed location $(x_k,y_k,0)$ on the ground, which is known {\it a-priori} by the UAV for designing the trajectory and transmit power allocation before it is dispatched for transmission. We consider a finite UAV mission/communication period $\mathcal{T} \triangleq (0,T]$, with duration $T>0$. During this period, the UAV flies at a fixed altitude $H>0$ (e.g. due to some regulations on UAVs), and thus its time-varying location is denoted as $(x(t),y(t),H)$ at any time instant $t \in \mathcal{T}$. By denoting the maximum UAV speed as $V$ in meter/second (m/s), we then have the following constraint:
\vspace{-0.5em}
\begin{align}
\label{speed cons}
\sqrt {\dot{x} ^2 (t) + \dot{y} ^2 (t)} \le V,\forall t \in \mathcal{T},
\end{align}
where $\dot{x}(t)$ and $\dot{y}(t)$ denote the time-derivatives of $x(t)$ and $y(t)$, respectively.

In practice, the wireless channel from the UAV to each ground user is dominated by the LoS component. Therefore, we consider the free-space path-loss model as in prior works \cite{ZengZhangLim2016-2,Xu2017}, and denote the channel power gain from the UAV to user $k \in \mathcal{K}$ as $h_k(t) =\beta _0 d_k^{ - 2}(t)$, where $d_k(t)=\sqrt {(x(t)-x_k)^2+(y(t)-y_k)^2+H^2} $ denotes their distance at time instant $t \in \mathcal{T}$ and $\beta _0$ denotes the channel power gain at a reference distance of $d_0=1~\mathrm{m}$.

Let $s(t)$ denote the common information signal transmitted by the UAV at time instant $t \in \mathcal{T}$. Then user $k$'s received signal is expressed as
\vspace{-0.5em}
\begin{align}
u_k(t)=\sqrt{h_k(t)} s(t)+v_k(t),\forall k \in \mathcal{K}, t \in \mathcal{T},
\end{align}
where $v_k(t)$ denotes the additive white Gaussian noise (AWGN) at the receiver of user $k$. Let $p(t)=\mathbb{E}(|s(t)|^2)$ denote the transmit power of the UAV at time instant $t$, where $\mathbb{E}(\cdot)$ denotes the statistical expectation. Suppose that the UAV has a maximum average transmit power constraint $P_{\text{ave}}$. We thus have
\vspace{-0.5em}
\begin{align}
\label{power cons}
\frac{1}{T} \int_0^T {p(t)} dt \le P_{\text{ave}}.
\end{align}

We are interested in characterizing the capacity of the UAV-enabled multicast channel, which is defined as the maximum achievable rate that can be simultaneously transmitted from the UAV to the $K$ ground users. Note that under a given UAV trajectory $\{(x(t),y(t))\}$ and transmit power allocation $\{p(t)\}$, the achievable rate from the UAV to each user $k$ over the mission/communication period $\mathcal{T}$ in bps/Hz is given by
\vspace{-0.5em}
\begin{align}
\hat{R}_k(\{x(t),y(t),p(t)\})=\frac{1}{T} \int_0^T R_k (x(t),y(t),p(t)) \text{d}t,
\end{align}
where
\vspace{-0.5em}
\begin{align}\label{Rkdefine}
& R_k\left(x(t),y(t),p(t)\right)  = \log_2 \left( 1 + \frac {p(t) h_k(t)} {\sigma ^2} \right)\nonumber\\
& = \log_2 \left( 1 + \frac {\gamma_0 p(t)} { \left( x(t)-x_k \right)^2 + \left( y(t)-y_k \right) ^2 + H ^2 } \right),
\end{align}
with $\gamma_0=\beta_0/\sigma ^2$ denoting the reference signal-to-noise ratio (SNR). Accordingly, the achievable rate of the UAV-enabled multicast channel is expressed as
\vspace{-0.5em}
\begin{align}\label{achi rate}
\hat{R}(\{x(t),y(t),p(t)\})&=\min_{k \in \mathcal{K}} \hat{R}_k (\{x(t),y(t),p(t)\})\nonumber\\
=&\min_{k \in \mathcal{K}}~\frac{1}{T}\int_0^T R_k (x(t),y(t),p(t)) \text{d}t.
\end{align}
In order to practically achieve the rate in \eqref{achi rate}, the UAV needs to use a sufficiently long code that spans over the whole mission/communication period. Note that a similar code has been used in the stochastic transmit beamforming in multi-antenna multicast channels in \cite{Wu2013}, in which the code spans over different fading states to achieve higher rates.

Our objective is to characterize the capacity of the UAV-enabled multicast channel, which corresponds to maximizing the achievable rate $\hat{R}(\{x(t),y(t),p(t)\})$ defined in \eqref{achi rate} \cite{Jindal2006}, via jointly optimizing the trajectory $\left\{ x(t),y(t) \right\}$ and the transmit power allocation $\left\{ p(t)\right\}$  of the UAV, subject to the maximum speed constraint in \eqref{speed cons} and the average transmit power constraint in \eqref{power cons}. Mathematically, we formulate the capacity optimization problem as
\vspace{-0.5em}
\begin{align}
\text{(P1)}: \max_{\left\{ x(t),y(t),p(t) \right\}} &~ \min_{k \in \mathcal{K}} ~ \frac{1}{T}\int_0^T {R_k\left(x(t),y(t),p(t)\right)}\text{d}t \nonumber\\
\label{power cons>0}
 \mathrm{s.t.} ~~~~~&~ p(t) \ge 0, \forall t \in \mathcal{T}\\
&~\eqref{speed cons}~\text{and}~\eqref{power cons}.\nonumber
\end{align}
Note that in problem (P1), the objective function is non-convex in general. Therefore, problem (P1) is non-convex and thus difficult to be solved optimally. To tackle this problem, in Section III we first present the optimal solution to a relaxed problem of (P1) given in the following, by ignoring the maximum UAV speed constraint \eqref{speed cons}.
\vspace{-0.5em}
\begin{align}
\text{(P2)}: \max_{\left\{ x(t),y(t),p(t) \right\}} & \min_{k \in \mathcal{K}} ~ \frac{1}{T}\int_0^T {R_k\left(x(t),y(t),p(t)\right)}\text{d}t \nonumber\\
 \mathrm{s.t.}~~~~ & \eqref{power cons}~\text{and}~\eqref{power cons>0}.\nonumber
\end{align}
Note that the constraint in \eqref{speed cons} can be approximately ignored in practice when the maximum UAV speed $V$ and/or the communication duration $T$ become sufficiently large. In Section IV, we present an efficient solution to problem (P1) based on the optimal solution to (P2).

\section{Optimal Solution to Problem (P2)}
Problem (P2) can be equivalently expressed as the following problem by introducing an auxiliary variable $\eta$.
\vspace{-0.5em}
\begin{align}
\text{(P2.1)}: &\max_{\left\{x(t),y(t),p(t)\right\},\eta}  ~ \eta\nonumber\\
\label{p2.1 cons1}
&~~~\mathrm{s.t.} ~  \int_0^T {R_k\left(x(t),y(t),p(t)\right)}\text{d}t \ge T\eta,\forall k \in \mathcal{K}\\
&~~~~~~~~\eqref{power cons}~\text{and}~\eqref{power cons>0}.\nonumber
\end{align}
Although (P2.1) is still a non-convex problem, it satisfies the so-called time-sharing condition in \cite{YuLui2006}. Therefore, strong duality holds between (P2.1) and its Lagrange dual problem. As a result, we can optimally solve (P2.1) by using the Lagrange dual method \cite{Boyd:Book}.

Let $\lambda_k, k\in\mathcal{K},$ and $\mu$ denote the non-negative Lagrange multipliers associated with the $k$-th constraint in \eqref{p2.1 cons1} and the constraint \eqref{power cons}, respectively. The partial Lagrangian of problem (P2.1) is thus given by
\vspace{-0.5em}
\begin{align}
& \mathcal{L} \left( \left\{x(t),y(t),p(t) \right\},\eta,\{\lambda_k\},\mu \right)\nonumber\\
& =\sum_{k \in \mathcal{K}} \lambda_k  \int_0^T R_k\left(x(t),y(t),p(t)\right)\text{d}t \nonumber\\
& ~ ~ +\bigg(1-T\sum_{k \in \mathcal{K}}\lambda_k \bigg) \eta - \mu\bigg(\frac{1}{T} \int_0^T {p(t)} \text{d}t-P_{\text{ave}}\bigg).
\end{align}
Accordingly, the dual function of (P2.1) is expressed as
\vspace{-0.5em}
\begin{align}
\label{dual func}
f \left( \{ \lambda_k \},\mu \right) = \max_{\left\{ x(t),y(t),p(t) \right\},\eta } &\mathcal{L} \left( \left\{x(t),y(t),p(t) \right\},\eta,\{\lambda_k\},\mu \right)\nonumber\\
\mathrm{s.t.} ~~~~& \eqref{power cons>0}.
\end{align}

\begin{lemma}\label{lemma3}
In order for $f\left( \{ \lambda_k \},\mu \right)$ to be bounded from above (i.e., $f\left( \{ \lambda_k \},\mu \right)<\infty$), it must hold that $\sum_{k \in \mathcal{K}} \lambda_k = 1/T$.
\end{lemma}
\begin{IEEEproof}
Suppose that $\sum_{k \in \mathcal{K}} \lambda_k > 1/T$ (or $\sum_{k \in \mathcal{K}} \lambda_k < 1/T$). Then by setting $\eta  \to  - \infty $ (or $\eta  \to   \infty $), we have $f\left( \{ \lambda_k \} ,\mu\right) \to \infty $. Therefore, this lemma is proved.
\end{IEEEproof}

Based on Lemma \ref{lemma3}, the dual problem of (P2.1) is given by
\vspace{-0.5em}
\begin{align}
\text{(D2.1)}: \min_{\{ \lambda_k \},\mu} ~ & f \left( \{ \lambda_k \},\mu \right) \nonumber\\
\label{the gather of lambda1}
\mathrm{s.t.} ~ & \sum_{k \in \mathcal{K}} \lambda_k = 1/T\\
\label{the gather of lambda2}
& \lambda_k \ge 0,\forall k \in \mathcal{K}\\
\label{the gather of mu}
& \mu \ge 0.
\end{align}

As strong duality holds between problem (P2.1) and its dual problem (D2.1), we can solve (P2.1) by equivalently solving (D2.1). Let the feasible set of $\{ \lambda_k \}$ and $\mu$ specified by \eqref{the gather of lambda1}, \eqref{the gather of lambda2}, and \eqref{the gather of mu} as $\chi$. In the following, we first solve problem \eqref{dual func} to obtain $f\left( \{ \lambda_k \},\mu \right)$ under any given $\big(\{ \lambda_k \},\mu\big) \in \chi$, then solve (D2.1) to find the optimal $\{ \lambda_k \}$ and $\mu$ that minimize $f\left( \{ \lambda_k \},\mu \right)$, and finally construct the optimal primal solution to (P2.1) and thus solve (P2).

\subsubsection{Obtaining $f\left( \{ \lambda_k \},\mu  \right)$ by Solving Problem \eqref{dual func}}
\label{step1}
For any given $\big(\{ \lambda_k \},\mu\big) \in \chi$, problem \eqref{dual func} can be decomposed into the following subproblems.
\vspace{-0.5em}
\begin{align}
\label{subpro1}
\max_{\eta} \left( 1 - T\sum_{k \in \mathcal{K}} \lambda_k  \right)\eta,
\end{align}
\vspace{-0.5em}
\begin{align}
\label{subpro2}
\max_{x(t),y(t),p(t)} &\sum_{k \in \mathcal{K}} \lambda_k R_k\left(x(t),y(t),p(t)\right)  - \mu p(t),~\forall t\in\mathcal T\nonumber\\
\text{s.t.}~&p(t) \ge 0 .
\end{align}
Here, \eqref{subpro2} consists of an infinite number of subproblems, each corresponding to one time instant $t \in \mathcal{T}$.  Let $\eta_{ \{ \lambda_k \},\mu }^*$ denote the optimal solution to problem \eqref{subpro1}, and $x_{ \{ \lambda_k \},\mu }^*(t)$, $y_{ \{ \lambda_k \},\mu }^*(t)$, and $p_{ \{ \lambda_k \},\mu }^*(t)$ denote that to problem \eqref{subpro2}.

As for problem \eqref{subpro1}, since $\sum_{k \in \mathcal{K}} \lambda_k = 1/T$ holds for any given $\big(\{ \lambda_k \},\mu\big) \in \chi$, the objective value is always zero. Thus, we can choose any arbitrary real number as the optimal solution $\eta_{ \{ \lambda_k \},\mu }^*$.

As for \eqref{subpro2}, note that all the subproblems are identical for different time indices $t$'s. Therefore, we can drop the time index $t$ and rewrite problem \eqref{subpro2} as
\vspace{-0.5em}
\begin{align}
\label{drop t problem}
& \max_{x,y,p\ge0} ~ \psi(x,y,p) \triangleq \nonumber\\
& \sum_{k \in \mathcal{K}} \lambda_k \log_2 \left( 1 + \frac{\gamma_0 p}{\left( x - x_k \right)^2 + \left( y - y_k \right)^2 + H^2} \right)-\mu p.
\end{align}
However, problem \eqref{drop t problem} is still non-convex, and thus is difficult to solve. Fortunately, there are only three variables in \eqref{drop t problem}; and for any given $x$ and $y$, $\psi(x,y,p)$ is a concave function with respect to $p\ge0$. Therefore, we can first use a simple bisection search to obtain the optimal $p$ under any given $x$ and $y$, denoted as $p^*(x,y)$, and then adopt a two-dimensional (2D) exhaustive search to find the optimal solution of $x$ and $y$ to problem \eqref{drop t problem}, denoted as $x_{\{ \lambda_k \},\mu}^*$ and $y_{\{ \lambda_k \},\mu}^*$. Note that we must have $(x_{\{ \lambda_k \},\mu}^* , y_{\{ \lambda_k \},\mu}^*) \in [ \underline x ,\overline x ] \times [ \underline y ,\overline y ]$, where
\vspace{-0.5em}
\begin{align}
\underline x  = \min _{k \in \mathcal{K}} x_k,\overline x  = \max _{k \in \mathcal{K}} x_k,\underline y  = \min _{k \in \mathcal{K}} y_k,\overline y  = \max _{k \in \mathcal{K}} y_k,
\end{align}
since otherwise, the UAV can always move its location into this box region to improve the objective value in \eqref{drop t problem}. Therefore, the 2D exhaustive search only needs to be adopted within $[ \underline x ,\overline x ] \times [ \underline y ,\overline y ]$ to find $x_{\{ \lambda_k \},\mu}^*$ and $y_{\{ \lambda_k \},\mu}^*$, i.e.,
\vspace{-0.5em}
\begin{align}
(x_{\{ \lambda_k \},\mu}^*,y_{\{ \lambda_k \},\mu}^*)=\argmax_{x \in [ \underline x ,\overline x ],y \in [ \underline y ,\overline y ] }~ \psi(x,y,p^*(x,y)).
\end{align}
As a result, the optimal solution to problem \eqref{drop t problem} is obtained as $x_{\{ \lambda_k \},\mu}^*$, $y_{\{ \lambda_k \},\mu}^*$, and $p^*(x_{\{ \lambda_k \},\mu}^*,y_{\{ \lambda_k \},\mu}^*)$.

Based on the optimal solution to problem \eqref{drop t problem}, we can obtain the optimal solution to problem \eqref{subpro2} as
\vspace{-0.5em}
\begin{align}
&x_{\{ \lambda_k \},\mu}^*(t) = x_{\{ \lambda_k \},\mu}^*,y_{\{ \lambda_k \},\mu}^*(t) = y_{\{ \lambda_k \},\mu}^*,\nonumber\\
&p_{\{ \lambda_k \},\mu}^*(t) = p^*(x_{\{ \lambda_k \},\mu}^*,y_{\{ \lambda_k \},\mu}^*),~\forall t \in \mathcal{T}.
\end{align}
Note that the optimal solution of $x_{\{ \lambda_k \},\mu}^*$, $y_{\{ \lambda_k \},\mu}^*$, and $p_{\{ \lambda_k \},\mu}^*$ to problem \eqref{drop t problem} is generally non-unique (as will be shown in numerical results). In this case, we can arbitrarily choose any one of them for the purpose of obtaining the dual function $f\left( \{ \lambda_k \},\mu \right)$.

\subsubsection{Finding Optimal Dual Solution to (D2.1)}
\label{step2}
Next, we solve the dual problem (D2.1) to find the optimal $\{ \lambda_k \}$ and $\mu$. Note that the dual function $f\left( \{ \lambda_k \},\mu \right)$ is always convex but in general non-differentiable \cite{Boyd:Book}. Therefore, we can apply the ellipsoid method to solve the dual problem (D2.1). Note that the subgradient of the dual function $f\left( \{ \lambda_k \},\mu \right)$ is
\vspace{-0.5em}
\begin{align}
&s_0(\lambda_1,...,\lambda_K,\mu)=\nonumber\\
&\bigg[TR_1\big(x_{ \{ \lambda_k \},\mu }^*,y_{ \{ \lambda_k \},\mu }^*,p^*(x_{\{ \lambda_k \},\mu}^*,y_{\{ \lambda_k \},\mu}^*)\big),...,\bigg.\nonumber\\
&~~TR_K\big(x_{ \{ \lambda_k \},\mu }^*,y_{ \{ \lambda_k \},\mu }^*,p^*(x_{\{ \lambda_k \},\mu}^*,y_{\{ \lambda_k \},\mu}^*)\big),\nonumber\\
&~~~~\bigg.{p^*(x_{\{ \lambda_k \},\mu}^*,y_{\{ \lambda_k \},\mu}^*)} -P_{\text{ave}}\bigg],
\end{align}
where $\eta_{ \{ \lambda_k \},\mu }^*=0$ is chosen for simplicity.
Also note that the equality constraint in \eqref{the gather of lambda1} can be viewed as two inequality constraints $\sum_{k \in \mathcal{K}} \lambda_k -1 \le 0$ and $-\sum_{k \in \mathcal{K}} \lambda_k +1 \le 0$, whose subgradients are respectively given by $
[1,\ldots,1,0]$ and $[-1,\ldots,-1,0]$. Let the obtained optimal solution to (D2.1) be denoted by $\{ \lambda_k^\star \}$ and $\mu^\star$.

\subsubsection{Constructing Optimal Primal Solution to (P2.1)}
With $\{ \lambda_k^\star \}$ and $\mu^\star$ at hand, we then obtain the optimal primal solution to (P2.1), denoted as $\left\{ x^\star(t) \right\}$, $\left\{ y^\star(t) \right\}$, $\left\{ p^\star(t) \right\}$, and ${\eta ^\star}$.

Note that under the optimal dual variables $\{ \lambda_k^\star \}$ and $\mu^\star$, if the optimal solution of $\{x^*_{\{\lambda_k^\star\},\mu^\star}(t)\}$, $\{y^*_{\{\lambda_k^\star\},\mu^\star}(t)\}$, $\{p^*_{\{\lambda_k^\star\},\mu^\star}(t)\}$ to problem \eqref{dual func} is unique, then it is also the optimal primal solution to (P2.1). However, if the optimal solution to problem \eqref{dual func} is not unique, then we need to reconstruct the optimal primal solution to (P2.1) by time-sharing them, as shown in the following.

In particular, suppose that under $\{ \lambda_k^\star \}$ and $\mu^\star$,  problem \eqref{drop t problem} has $\Gamma \ge 1$ optimal solutions, denoted by $\{x_\phi^*,y_\phi^*,p_\phi^* \}_{\phi=1}^\Gamma$, i.e., the UAV needs to hover at $\Gamma$ optimal locations $(x_\phi^*,y_\phi^*,H),\phi \in \{1,\ldots,\Gamma\}$, each with an optimal transmit power $p_\phi^*$. To obtain the optimal primal solution to (P2.1) in this case, we need to time-share among the $\Gamma$ solutions, i.e., the UAV should hover at each of the  $\Gamma$ optimal locations for a certain duration that needs to be optimized. Let $t_\phi\ge0$ denote the hovering duration at $(x_\phi^*,y_\phi^*,H), \phi \in \{1,...,\Gamma \}$, where $\sum_{\phi=1}^{\Gamma} t_\phi=T$. Accordingly, finding the optimal hovering durations $\{t_\phi\}$ corresponds to solving the following problem.
\vspace{-0.5em}
\begin{align}
\label{hoverdurapro}
\max_{ \{ t_\phi \ge 0\},\eta} & ~ \eta\nonumber\\
\mathrm{s.t.} & ~ \sum_{\phi=1}^\Gamma t_\phi R_k \left( x_\phi^*,y_\phi^*,p_\phi^* \right) \ge T\eta,\forall k \in \mathcal{K}\nonumber\\
& \sum_{\phi=1}^\Gamma t_\phi = T.
\end{align}
Problem \eqref{hoverdurapro} is a linear programming (LP), which can be solved efficiently by using standard convex optimization techniques \cite{Boyd:Book}. Let $\eta^\star$ and $\{t_\phi^\star\}$ denote the optimal solution to problem \eqref{hoverdurapro}. Accordingly, we partition the whole communication period $\mathcal T$ into $\Gamma$ sub-periods, denoted by $\mathcal T_1,\ldots,\mathcal T_\Gamma$, where $\mathcal T_\phi = (\sum_{i=1}^{\phi-1} t_i^\star, \sum_{i=1}^{\phi} t_i^\star]$, $\forall \phi \in \{1,\ldots,\Gamma\}$. In this case, the optimal value (or the capacity of the UAV-enabled multicast channel) is given by $\eta^\star$, and the optimal trajectory and power allocation solution to the primal problem (P2.1) (also (P2)) is given as
\begin{align}
\label{optimal tra}
x^\star(t)=x_\phi^*,y^\star(t)=y_\phi^*,p^\star(t)=p_\phi^*,\forall t \in \mathcal{T}_\phi,\phi \in \{1,...,\Gamma\}.
\end{align}

\begin{remark}\label{remark:1}
The optimal solution in \eqref{optimal tra} to problem (P2) reveals that to achieve the capacity of this multicast channel (with the maximum UAV speed constraint ignored), the UAV should hover above a finite number of ground locations, with the optimal hovering duration and transmit power at each location. This can be intuitively explained as follows. By hovering at different locations each of which is closer to a different subset of users in general, the UAV can have better wireless channels to them and thus improve the minimum average rate of all users. This helps overcome the bottleneck user issue in conventional multicast channels with fixed transmitters and hence significantly improves the multicast channel capacity. We refer to the above optimal solution as a {\it multi-location-hovering} solution.
\end{remark}

\section{Proposed Solution to Problem (P1)}\label{sec:proposed:P1}

In this section, we consider problem (P1) with the maximum UAV speed constraint in \eqref{speed cons} considered. First, we propose a successive hover-and-fly trajectory design based on the optimal multi-location-hovering solution to problem (P2), and then jointly optimize the corresponding hovering durations and the transmit power allocation for the UAV.

\subsection{Succeesive Hover-and-Fly Trajectory Design}
Recall that at the optimal solution to problem (P2), the UAV needs to hover at $\Gamma$ hovering locations, denoted by $\{(x^*_\phi,y^*_\phi,H)\}_{\phi=1}^\Gamma$. Motivated by this multi-location-hovering solution, we propose a successive hover-and-fly trajectory design, in which the UAV successively visits these $\Gamma$ hovering locations. To maximize the hovering time for efficient information multicasting, the UAV should fly at the maximum speed $V$ between consecutive hovering locations, and minimize the total flying distance, provided that each hovering location is visited once. The flying distance minimization problem is similar to the travelling salesman problem (TSP), with only the following difference: in the TSP, the salesman (or the UAV of our interest here) needs to return to the initial location after visiting all the locations, but in our flying distance minimization problem, the UAV does not need to return to the initial location. Despite this difference, we show that the min-flying-distance problem can be solved via equivalently solving the following TSP \cite{Xu2017}. We construct a new TSP by adding a dummy location as the initial/final location of the trajectory, and letting the distances between the newly added location and the $\Gamma$ hovering locations be zero. By solving this new TSP with $\Gamma +1$ locations and then dropping the dummy location, we can find the optimal trajectory for the flying distance minimization problem. With the obtained solution, we use a permutation $\pi$ to denote the ordering of the visited hovering locations, i.e., the $\pi(1)$-th location is visited first, followed by the $\pi(2)$-th, $\pi(3)$-th, etc., until the $\pi(\Gamma)$-th hovering location is visited last. In this case, we define the distance between the $i$-th and the $j$-th hovering locations as
\vspace{-0.5em}
\begin{align}
d_{ij}=\sqrt{(x_i-x_j)^2+(y_i-y_j)^2}, \forall  i,j\in\{1,\ldots,\Gamma\},i\neq j.
\end{align}
Then the total flying distance is given as
\vspace{-0.5em}
\begin{align}
D=\sum_{\phi=1}^{\Gamma-1}~d_{\pi(\phi)\pi(\phi+1)}.
\end{align}
Accordingly, the flying time from the $\pi(\phi)$-th to the $\pi(\phi+1)$-th hovering locations is $\tau_\phi^{\text{fly}}=d_{\pi(\phi)\pi(\phi+1)}/V$, and the total flying time is $T_\text{fly}=D/V=\sum_{\phi=1}^{\Gamma-1} \tau_\phi^{\text{fly}}$. We denote the obtained flying trajectory as $\{ \hat{x}(t),\hat{y}(t)\}_{t=0}^{T_\text{fly}}$. Note that in order for the UAV to have enough time to fly along the obtained flying trajectory to visit the $\Gamma$ hovering locations, in this paper we focus on the scenario when the mission/communication duration $T$ is larger than the total flying time $T_\text{fly}$ (i.e., $T \ge T_\text{fly}$).{\footnote{The successive hover-and-fly trajectory can also be extended in the scenario when $T < T_\text{fly}$, similarly as in \cite{Xu2017}. Due to the space limitation, we omit the extension and leave it for future work.}}

In order to obtain the complete successive hover-and-fly trajectory, we denote $\tau_\phi^{\text{hover}} \ge 0$ as the hovering duration at the $\pi(\phi)$-th hovering location, where $\sum_{\phi=1}^{\Gamma} \tau_\phi^{\text{hover}} = T - T_\text{fly}$. Note that $\tau_\phi^{\text{hover}}$'s are a set of variables that should be optimized later. In this case, the successive hover-and-fly trajectory can be expressed as follows by dividing the mission/communication period $\mathcal T$ into $2\Gamma -1$ sub-periods, denoted by $\hat{\mathcal T}_1,\ldots,\hat{\mathcal T}_{2\Gamma -1}$. The odd and even sub-periods are defined explicitly as follows for hovering and flying, respectively.
\vspace{-0.5em}
\begin{align*}
&\hat{\mathcal{T}}_{2\phi - 1} \triangleq \\
&\bigg(\sum_{i=1}^{\phi-1}(\tau_i^{\text{hover}}+\tau_i^{\text{fly}}), \sum_{i=1}^{\phi-1}(\tau_i^{\text{hover}}+\tau_i^{\text{fly}})+\tau_\phi^{\text{hover}}\bigg],\forall \phi\in\{1,\ldots, \Gamma\}\\
&\hat{\mathcal{T}}_{2\phi} \triangleq \\
&\bigg(\sum_{i=1}^{\phi-1}(\tau_i^{\text{hover}}+\tau_i^{\text{fly}})+\tau_\phi^{\text{hover}}, \sum_{i=1}^{\phi}(\tau_i^{\text{hover}}+\tau_i^{\text{fly}})\bigg],\forall \phi\in\{1,\ldots, \Gamma-1\}
\end{align*}
Accordingly, within each odd sub-period $2\phi - 1$, the UAV should hover at the $\pi(\phi)$-th hovering location $(x_{\pi(\phi)}, y_{\pi(\phi)},H)$, i.e.,
\vspace{-0.5em}
\begin{align}
x(t) = x_{\pi(\phi)},~y(t) =  y_{\pi(\phi)}, \label{eqn:trajectory:1}
\end{align}
$\forall t\in \hat{\mathcal{T}}_{2\phi - 1}, \phi\in\{1,\ldots, \Gamma\}$. During each even sub-period $2\phi$, the UAV should fly from the $\pi(\phi)$-th hovering location to the $\pi(\phi+1)$-th hovering location at the maximum speed $V$, with time-varying location being
\vspace{-0.5em}
\begin{align}
x(t) = \hat{x}\bigg(t-\sum_{i=1}^{\phi} \tau_i^{\text{hover}}\bigg),~y(t) = \hat{y}\bigg(t-\sum_{i=1}^{\phi} \tau_i^{\text{hover}}\bigg),\label{eqn:trajectory:2}
\end{align}
$\forall t\in \hat{\mathcal{T}}_{2\phi},\phi\in\{1,\ldots, \Gamma-1\}$.

\subsection{Joint Optimization of Hovering Durations and Power Allocation}
Under the successive hover-and-fly trajectory, we need to decide the hovering durations $\{\tau_\phi^{\text{hover}}\}$ as well as the transmit power allocation $\{p(t)\}$ for the UAV. We first discretize the even sub-periods for flying with duration $T_{\text{fly}}$ into $N$ time slots with equal durations. The number of time slots $N$ is chosen to be sufficiently large, such that the duration of each time slot $\Delta_t$ is sufficiently small, during which the location and the power allocation of the UAV are approximately unchanged. Then based on the successive hover-and-fly trajectory, we can obtain the location of the UAV at time slot $j \in \{1,...,N\}$ as $\left( x^\text{fly}[j],y^\text{fly}[j],H \right)$, where $x^\text{fly}[j]=\hat{x}(j\Delta_t),y^\text{fly}[j]=\hat{y}(j\Delta_t)$. Let $p_\phi^{\text{hover}},\forall \phi \in \{1,...,\Gamma \}$ denote the power allocation when the UAV hovers above the $\pi(\phi)$-th hovering location, and $p^\text{fly}[j],\forall j \in \{1,...,N\}$ denote the power allocation when the UAV flies at time slot $j$. Accordingly, the achievable rate of user $k$ is
\vspace{-0.5em}
\begin{align}
&\overline{R}_k(\{\tau_\phi^{\text{hover}},p_\phi^{\text{hover}}\},\{p^\text{fly}[j]\})\nonumber\\
=&\frac{1}{T}\bigg(\Delta_t\sum_{j=1}^{N} R_k (x^\text{fly}[j],y^\text{fly}[j],p^\text{fly}[j])\nonumber\\
&~~+\sum_{\phi=1}^{\Gamma} \tau_\phi^{\text{hover}} R_k (x^*_{\pi(\phi)},y^*_{\pi(\phi)},p_\phi^{\text{hover}})\bigg),
\end{align}
in which $R_k(\cdot,\cdot,\cdot)$ is given in \eqref{Rkdefine}.

Based on the discretization and by introducing an auxiliary variable $\eta$, problem (P1) under the successive hover-and-fly trajectory can be reformulated as follows by jointly optimizing the hovering durations and transmit power allocation.
\vspace{-0.5em}
\begin{align}\label{opti dura}
&\max_{ \{\tau_\phi^{\text{hover}},p_\phi^{\text{hover}}\},\{p^\text{fly}[j]\}, \eta}  \eta\\
&~~~~\mathrm{s.t.} ~ \overline{R}_k(\{\tau_\phi^{\text{hover}},p_\phi^{\text{hover}}\},\{p^\text{fly}[j]\}) \ge \eta,\forall k \in \mathcal{K}\nonumber\\
&~~~~~~~~~ \sum_{\phi=1}^{\Gamma} \tau_\phi^{\text{hover}} p_\phi^{\text{hover}} + \Delta_t\sum_{j=1}^{N} p^\text{fly}[j] \le T P_{\text{ave}}\nonumber\\
&~~~~~~~~~ \sum_{\phi=1}^{\Gamma} \tau_\phi^{\text{hover}}=T-T_\text{fly}\label{opti dura:con1}\\
&~~~~~~~~~ \tau_\phi^{\text{hover}} \ge 0,p_\phi^{\text{hover}}\ge 0, \forall \phi \in \{1,...,\Gamma \}\label{opti dura:con2}\\
&~~~~~~~~~ p^\text{fly}[j] \ge 0,\forall j\in\{1,\ldots,N\}.\label{opti dura:con3}
\end{align}
However, problem \eqref{opti dura} is non-convex, due to the coupling between the hovering duration $\tau_\phi^{\text{hover}}$ and the power allocation $p_\phi^{\text{hover}}$. Nevertheless, by introducing auxiliary variables
\vspace{-0.5em}
\begin{align}\label{change_variable}
E_\phi=\tau_\phi^{\text{hover}} p_\phi^{\text{hover}},\forall \phi \in \{1,...,\Gamma \},
\end{align}
problem \eqref{opti dura} can be re-expressed as the following convex optimization problem.
\vspace{-0.5em}
\begin{align}\label{opti dura1}
&\max_{ \{\tau_\phi^{\text{hover}},E_\phi\},\{p^\text{fly}[j]\}, \eta} ~ \eta\\
&~~~~ \mathrm{s.t.} ~ \frac{1}{T} \bigg( \sum_{\phi=1}^{\Gamma} \tau_\phi^{\text{hover}} \log_2\bigg(1+\frac{\alpha_{k,\phi} E_\phi}{\tau_\phi^{\text{hover}}}\bigg)\nonumber\\
&~~~~~~~~~~~~+\Delta_t\sum_{j=1}^{N} R_k (x[j],y[j],p^\text{fly}[j])\bigg)\ge \eta,\forall k \in \mathcal{K}\nonumber\\
&~~~~~~~~~ \sum_{\phi=1}^{\Gamma} E_\phi + \Delta_t\sum_{j=1}^{N} p^\text{fly}[j] \le T P_{\text{ave}}\nonumber\\
&~~~~~~~~~ \tau_\phi^{\text{hover}} \ge 0,E_\phi\ge 0, \forall \phi \in \{1,...,\Gamma \}\nonumber\\
&~~~~~~~~~\eqref{opti dura:con1}~{\text{and}}~\eqref{opti dura:con3},\nonumber
\end{align}
where $\alpha_{k,\phi}=\phi_0/((x_k-x_{\pi(\phi)})^2+(y_k-y_{\pi(\phi)})^2+H^2), \forall k \in \mathcal{K},\phi \in [1,...,\Gamma],$ are constants. Let $\{\tau_\phi^{\text{hover}{\star\star}},E^{\star\star}_\phi\},\{p^{\text{fly}\star\star}[j]\},$ and $\eta^{\star\star}$ denote the optimal solution to problem \eqref{opti dura1}. By using them together with \eqref{change_variable}, we can obtain the optimal solution to problem \eqref{opti dura} as $\{\tau_\phi^{\text{hover}{\star\star}},p^{\star\star}_\phi\},\{p^{\text{fly}\star\star}[j]\},$ and $\eta^{\star\star}$. By combining them with the trajectory in \eqref{eqn:trajectory:1} and \eqref{eqn:trajectory:2}, the joint trajectory design and power allocation solution to (P1) is finally obtained.


\begin{remark}\label{remark:2}
Note that when the mission/communication period $T$ becomes large, our proposed design with joint successive hover-and-fly trajectory and transmit power allocation is asymptotically optimal for problem (P1). This is due to the fact that in this case, the flying time among hovering locations becomes negligible, and thus the achievable rate by the successive hover-and-fly trajectory with optimal power allocation approaches the optimal objective value of (P2) by the multi-location-hovering solution, which serves as an upper bound for the optimal value of (P1).
\end{remark}

\section{Numerical Results}
In this section, we present numerical results to validate the performance of our proposed joint trajectory and power optimization design, as compared to the following two benchmark schemes.

\subsubsection{Static hovering}
The UAV hovers at one fixed location over the whole mission/communication period. We need to optimize the fixed location and power allocation of the UAV to maximize the achievable rate as follows.
\vspace{-0.5em}
\begin{align}\label{static}
\max_{x,y,\left\{p(t)\right\}} ~& \min_{\forall k \in \mathcal{K}}~\frac{1}{T}\int_0^T {R_k\left(x,y,p(t)\right)}\text{d}t\nonumber\\
\mathrm{s.t.} ~ & \eqref{power cons}~\text{and}~\eqref{power cons>0}.
\end{align}
Under any given $x$ and $y$, $R_k\left(x,y,p(t)\right)$ is a concave function with respect to $p(t)$. Based on the Jensen's inequality, we must have $p(t)=P_{\text{ave}},\forall t \in \mathcal{T}$, to maximize $\frac{1}{T}\int_0^T {R_k\left(x,y,p(t)\right)}\text{d}t$ for any $k\in\mathcal K$. Therefore, it follows that $p(t)=P_{\text{ave}},\forall t \in \mathcal{T}$, at the optimal solution to problem \eqref{static}. As a result, we only need to use a 2D exhaustive search to obtain the optimal fixed hovering location of the UAV, i.e.,
\vspace{-0.5em}
\begin{align}\label{benchmark1}
(x^{\text{sta}},y^{\text{sta}})=\argmax_{x,y}~\min_{k \in \mathcal{K}}R_k\left( x,y,P_{\text{ave}}\right).
\end{align}

\subsubsection{Successive hover-and-fly trajectory with equal power allocation}

Similarly as in the proposed solution to (P1) in Section \ref{sec:proposed:P1}, the UAV successively visits the $\Gamma$ optimal hovering locations, with the trajectory given in \eqref{eqn:trajectory:1} and \eqref{eqn:trajectory:2}. As the equal power allocation is adopted, the optimal hovering durations at these hovering locations can be obtained by solving problem \eqref{opti dura} by setting $p_\phi^{\text{hover}} = P_{\text{ave}},\forall \phi \in\{1,\ldots,\Gamma\}$ and $p^\text{fly}[j] = P_{\text{ave}}, \forall j$.


In the simulation, we consider an example with $K=10$ users that are randomly distributed in an area of $1000~ \text{m} \times 1000~ \text{m}$, as shown in Fig.~1. The UAV flies at a fixed altitude of $H=100~\mathrm{m}$. The maximum average transmit power and maximum speed of the UAV are set as $P_{\text{ave}}=30~\mathrm{dBm}$ and $V=20~\mathrm{m/s}$, respectively. The noise power at the receiver of each user (including both thermal and background noise) is set as $\sigma^2=-50~\mathrm{dBm}$. The channel power gain at the reference distance of 1 m is set as $\beta_0=-30~\mathrm{dB}$.

\begin{figure}
\centering
 \epsfxsize=1\linewidth
    \includegraphics[width=6.7cm]{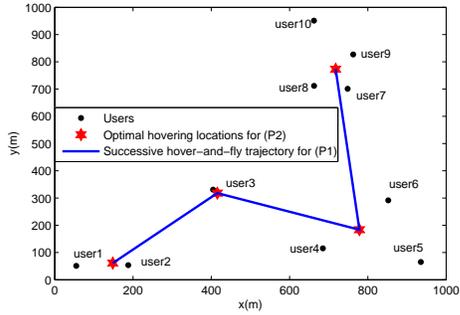}
\caption{Simulation example with $K=10$ ground users.}
\end{figure}

Fig. 1 shows the obtained optimal multi-location-hovering solution to (P2) without the maximum UAV speed constraint, and the successive hover-and-fly trajectory for (P1) with the maximum UAV speed constraint. It is observed that there are a total of $\Gamma=4$ optimal hovering locations for (P2) and intuitively, each hovering location is close to one or more ground users for efficient information multicasting.

\begin{figure}
\centering
 \epsfxsize=1\linewidth
    \includegraphics[width=6.7cm]{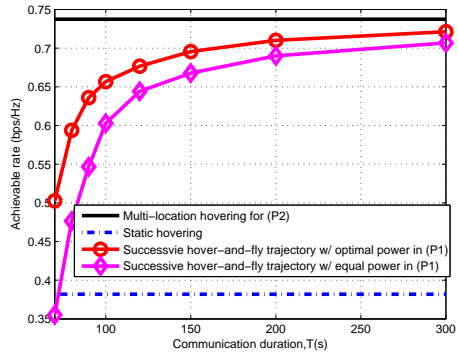}
\caption{Achievable rates of the UAV-enabled multicast channel, versus the communication duaration $T$.}\vspace{-0em}
\end{figure}

Fig. 2 shows the achievable rates of the UAV-enabled multicast channel in Fig. 1, versus the UAV communication duration $T$. It is observed that both the successive hover-and-fly trajectory designs with optimal and equal power allocation significantly outperform the static-hovering benchmark when $T \ge 80$~s, and the gain becomes more substantial when $T$ is larger. This validates the benefit of exploiting the UAV mobility in information multicasting. It is also observed that the successive hover-and-fly trajectory design with optimal power allocation achieves higher achievable rate than that with equal power allocation. This indicates the importance of power allocation for such a UAV-enabled multicast channel. Finally, the successive hover-and-fly trajectory design with optimal power allocation is observed to approach the performance upper bound by the multi-location-hovering solution to (P2) as $T$ increases. This is consistent with Remark \ref{remark:2}.

\section{Conclusion}
In this paper, we studied a UAV-enabled information multicasting system, where a UAV is dispatched as a mobile transmitter to deliver common information to a set of ground users over a given UAV mission/communication period. We characterized the capacity of this new UAV-enabled multicast channel by maximizing the minimum time-averaged achievable rates of all users via jointly optimizing the trajectory and the transmit power allocation of the UAV. To achieve the capacity, the UAV should use a sufficiently long code that spans over the whole mission/communication period. As the minimum-rate maximization problem is non-convex and difficult to solve, we first considered a relaxed problem without the maximum UAV speed constraint and obtained its optimal solution. Based on this optimal solution, we proposed an efficient design with successive hover-and-fly UAV trajectory and optimal power allocation for the case with the maximum UAV speed constraint considered. Numerical results validated the superior performance of our proposed design over benchmark schemes. It is our hope that this paper brings new insights on how to improve the communication rates of wireless multicast channels via exploiting the mobility of transmitters.

\section*{Acknowledgements}
This work was supported by the National Nature Foundation of China under Grant No.~61672484 and the National Natural Science Foundation of China under Grant No.~61628103.


\begin{thebibliography}{1}
\bibliographystyle{IEEEbib}

\bibitem{ZengZhangLim2016}
Y. Zeng, R. Zhang, and T. J. Lim, ``Wireless communications with unmanned aerial vehicles: Opportunities and challenges,'' {\it IEEE Commun. Mag.}, vol. 54, no. 5, pp. 36--42, May 2016.

\bibitem{LyuZengZhang2017}
J. Lyu, Y. Zeng, R. Zhang, and T. J. Lim, ``Placement optimization of UAV-mounted mobile base stations,'' {\it IEEE Commun. Lett.}, vol. 21, no. 3, pp. 604--607, Mar. 2017.

\bibitem{BorEl2016}
R. I. Bor-Yaliniz, A. El-Keyi, and H. Yanikomeroglu, ``Efficient 3-D placement of an aerial base station in next generation cellular networks,'' in {\it Proc. IEEE ICC}, pp. 1--5, May 2016.

\bibitem{Mozaffari2016}
M. Mozaffari, W. Saad, M. Bennis, and M. Debbah, ``Efficient deployment of multiple unmanned aerial vehicles for optimal wireless coverage,'' {\it IEEE Commun. Lett.}, vol. 20, no. 8, pp. 1647--1650, Aug. 2016.

\bibitem{Wu2017}
Q. Wu, Y. Zeng, and R. Zhang, ``Joint trajectory and communication design for multi-UAV enabled wireless networks,'' to appear in {\it IEEE Trans. Wireless Commun.}, 2018.

\bibitem{ZengZhangLim2016-2}
Y. Zeng, R. Zhang, and T. J. Lim, ``Throughput maximization for UAV-enabled mobile relaying systems,'' {\it IEEE Trans. Commun.}, vol. 64, no. 12, pp. 4983--4996, Dec. 2016.

\bibitem{WuXuZhang2017}
Q. Wu, J. Xu, R. Zhang, ``Capacity characterization of UAV-enabled two-User broadcast channel.'' [Online] Available: \url{https://arxiv.org/abs/1801.00443}.

\bibitem{Xu2017}
J. Xu, Y. Zeng, and R. Zhang, ``UAV-enabled wireless power transfer: Trajectory design and energy optimization.'' [Online] Available:
\url{https://arxiv.org/abs/1709.07590}.

\bibitem{XuZengZhang2017}
J. Xu, Y. Zeng, and R. Zhang, ``UAV-enabled wireless power transfer: Trajectory design and energy region characterization,'' in {\it Proc. IEEE Globecom Workshop}, 2017.

\bibitem{XieXuZhang2018}
L. Xie, J. Xu, and R. Zhang, ``Throughput maximization for UAV-enabled wireless powered communication networks,'' in {\it Proc. IEEE VTC2018-Spring}, 2018.

\bibitem{Jindal2006}
N. Jindal and Z.-Q. Luo, ``Capacity limits of multiple antenna multicast,'' in {\it Proc. IEEE Int. Symposium Inf. Theory (ISIT)}, pp. 1841--1845, 2006.

\bibitem{Sidiropoulos2006}
N. D. Sidiropoulos, T. N. Davidson, and Z.-Q. Luo, ``Transmit beamforming for physical-layer multicasting,'' {\it IEEE Trans. Signal Process.}, vol. 54, no. 6, pp. 2239--2251, Jun. 2006.

\bibitem{Xiang2013}
Z. Xiang, M. Tao, and X. Wang, ``Coordinated multicast beamforming in multicell networks,'' {\it IEEE Trans. Commun.}, vol. 12, no. 1, pp. 12--21, Jan. 2013.

\bibitem{Wu2013}
S. X. Wu, W.-K. Ma, and A. M.-C. So,  ``Physical-layer multicasting by stochastic transmit beamforming and Alamouti space-time coding,'' {\it IEEE Trans. Signal Process.}, vol. 61, no. 17, pp. 4230--4245, Sep. 2013.

\bibitem{YuLui2006}
W. Yu and R. Lui, ``Dual methods for nonconvex spectrum optimization of multicarrier systems,'' {\it IEEE Trans. Commun.}, vol. 54, no. 7, pp. 1310--1322, Jul. 2006.

\bibitem{ZengXu2017}
Y. Zeng, X. Xu, and R. Zhang, ``Trajectory design for completion time minimization in UAV-enabled multicasting.'' to appear in {\it IEEE Trans. Wireless Commun.}, 2018.

\bibitem{Boyd:Book}
S. Boyd and L. Vandenberghe, {\it Convex Optimization}. Cambridge, U.K.: Cambridge Univ. Press, 2004.

\end{thebibliography}
\end{document}